\documentclass[aps,reprint,showpacs]{revtex4-1}
\usepackage{graphicx} 
\usepackage{tabularx}
\usepackage{dcolumn} 
\usepackage{color}
\usepackage{bm}
\usepackage{amsmath}
\usepackage{amssymb}
\newcommand{\bi}[1]{\ensuremath{\boldsymbol{#1}}} 
\newcommand{\nn}{\nonumber}
 
\begin{document} 
 
\title{ 
Edge Current due to Majorana Fermions in Superfluid $^3$He A- and B-Phases
}

\author{Yasumasa Tsutsumi} 
\affiliation{Department of Physics, Okayama University, 
Okayama 700-8530, Japan} 
\author{Kazushige Machida} 
\affiliation{Department of Physics, Okayama University, 
Okayama 700-8530, Japan} 
\date{\today}

\begin{abstract} 
We propose a method utilizing edge current to observe Majorana fermions in the surface Andreev bound state for the superfluid $^3$He A- and B-phases.
The proposal is based on self-consistent analytic solutions of quasi-classical Green's function with an edge.
The local density of states and edge mass current in the A-phase or edge spin current in the B-phase can be obtained from these solutions.
The edge current carried by the Majorana fermions is partially cancelled by quasiparticles (QPs) in the continuum state outside the superfluid gap.
QPs contributing to the edge current in the continuum state are distributed in energy even away from the superfluid gap.
The effect of Majorana fermions emerges in the depletion of the edge current by temperature within a low-temperature range.
The observations that the reduction in the mass current is changed by $T^2$-power in the A-phase 
and the reduction in the spin current is changed by $T^3$-power in the B-phase establish the existence of Majorana fermions.
We also point out another possibility for observing Majorana fermions by controlling surface roughness.
\end{abstract} 
 
 
 
\maketitle 

\section{Introduction}

The superfluid $^3$He is firmly established as an anisotropic $p$-wave superfluid~\cite{vollhardt:book}.
At an edge of the superfluid $^3$He, the surface Andreev bound state emerges by the pairing symmetry
due to the bulk-surface correspondence.
Recently, the surface Andreev bound state has become a focus of concern from topological aspects.
The superfluid gap of the superfluid $^3$He among topological superfluids is closed
at the interface of a topologically trivial vacuum by a topological phase transition.
This is a topological aspect of the surface Andreev bound state~\cite{tanaka:2012}.
The topological features are quite different between the superfluid $^3$He A- and B-phases
because the A- (B-)phase is a time reversal symmetry broken (unbroken) state.
The A-phase is a chiral superfluid with the spontaneous edge mass current
while B-phase is a helical superfluid with the spontaneous edge spin current.
Quasiparticles (QPs) bound in the surface Andreev bound state behave as Majorana fermions
owing to the particle-hole symmetry~\cite{read:2000,schnyder:2008}.
The Majorana nature in the A- and B-phases is also distinctive.
Majorana fermions have a linear dispersion relation
forming a ``Majorana valley'' in the A-phase~\cite{tsutsumi:2010b,tsutsumi:2011b} 
and a ``Majorana cone'' in the B-phase~\cite{chung:2009,nagato:2009}.
This Majorana cone has been observed by recent experiments~\cite{murakawa:2009,murakawa:2011}.
However, there has been no firm evidence of the edge current that accompanies Majorana fermions.

For other candidates of the topological superconductor, e.g., 
Sr$_2$RuO$_4$~\cite{mackenzie:2003,kashiwaya:2011}, UPt$_3$~\cite{machida:2012,tsutsumi:cond2}, and Cu$_x$Bi$_2$Se$_3$~\cite{sasaki:2011,yamakage:2012},
the precise pairing symmetry has not yet been identified and has been under intense discussion.
Thus, there has been no concrete topological superconductors established 
as the superfluid $^3$He in the chiral A-phase and helical B-phase thus far.

In the two-dimensional (2D) Fermi surface model of the A-phase,
namely, the fully gapped chiral $p$-wave state as the proposed pairing symmetry for Sr$_2$RuO$_4$~\cite{mackenzie:2003},
the edge mass current is carried both by Majorana QPs within the superfluid gap
and by the QPs in the continuum state outside the gap~\cite{stone:2004,sauls:2011}.
These two kinds of QPs have different contributions to the edge mass current and angular momentum.
Stone and Roy~\cite{stone:2004} first proved that,
in the 2D A-phase under a uniform pair potential, the magnitude of the angular momentum by the edge mass current is $N\hbar/2$ at a zero temperature
($N$ is the total number of $^3$He atoms in the whole system).
The A-phase with the three-dimensional (3D) Fermi surface under the pair potential reflected on an edge
has the same magnitude of this angular momentum and the same contributions from these two kinds of QPs~\cite{tsutsumi:2012}.

It is also expected that Majorana QPs and QPs in the continuum state have different contributions to the edge spin current in the B-phase.
In this study, we investigate the roles of these two kinds of QPs in the edge spin current
with the density of states (DOS) for these QPs.
It turns out that the edge spin current from Majorana QPs is partially cancelled
by that from QPs in the continuum state similarly to the A-phase~\cite{tsutsumi:2012}.
Finally, the total spin current at a zero temperature is $J_s=-(\kappa/2\pi)(n\hbar/6)$,
where $\kappa$ is the quantum of circulation and $n$ is the density of $^3$He atoms.
The numerical coefficient of $n\hbar$ for the spin current is $2/3$ of that for the total mass current $J=-n\hbar/4$ in the A-phase
because QPs contributing to the spin current are in two of three spin states of QPs.
The temperature dependence of the spin current will also be shown.
The depletion of the spin current by temperature within a low-temperature range is proportional to $T^3$
because the spin current from Majorana QPs has a quadratic energy spectrum.

We base our arguments on the quasi-classical theory, which is 
valid for $\xi\gg k_F^{-1}$, well satisfied for the superfluid $^3$He
(coherent length $\xi\sim 10$--100 nm and Fermi wave number $k_F^{-1}\sim 0.1$ nm).
We find analytic solutions for the pair potential when the system has an edge.
These analytic solutions give useful and transparent information on various physical properties at the edge.
Moreover, these solutions are self-consistent solutions in the infinite cutoff energy limit,
namely, the ``weak-coupling limit".
Thus, the present comparative studies on both the A- and B-phases based on the same theoretical framework may help us understand the nature of Majorana fermions.

This paper is arranged as follows:
In \S\ref{sec:theory}, we formulate the quasi-classical theory on the basis of the quasi-classical Green's function, which gives quantitative information on QPs.
We discuss the edge mass current and angular momentum in the A-phase in \S\ref{sec:A}
in comparison with the 2D Fermi surface model.
We also touch on the so-called intrinsic angular momentum problem.
In \S\ref{sec:B}, we investigate the role of Majorana QPs in the edge spin current with the DOS in the B-phase.
We also show the temperature dependence of the spin current.
We devote the final section to the summary and discussion.
We demonstrate the self-consistency of our solutions in the weak-coupling limit in Appendix.

\section{Quasi-Classical Theory}
\label{sec:theory}

Microscopic information on the edge state is contained in the quasi-classical Green's function
$\widehat{g}(\bi{r},\bi{k},\omega_n)$.
The quasi-classical Green's function is calculated using the Eilenberger equation~\cite{eilenberger:1968} as
\begin{align}
-i\hbar\bi{v}(\bi{k})\cdot\bi{\nabla }\widehat{g}(\bi{r},\bi{k},\omega_n)
= \left[i\omega_n\widehat{\sigma }_z-\widehat{\Delta }(\bi{r},\bi{k}),
\widehat{g}(\bi{r},\bi{k},\omega_n) \right].
\label{Eilenberger eq}
\end{align}
In this paper, the ``ordinary hat" and ``wide hat" indicate the 2 $\times$ 2 matrix in spin space 
and the 4 $\times$ 4 matrix in particle-hole and spin spaces, respectively.
The quasi-classical Green's function is described in particle-hole space by
\begin{align}
\widehat{g}(\bi{r},\bi{k},\omega_n) = -i\pi
\begin{pmatrix}
\hat{g}(\bi{r},\bi{k},\omega_n) & i\hat{f}(\bi{r},\bi{k},\omega_n) \\
-i\underline{\hat{f}}(\bi{r},\bi{k},\omega_n) & -\underline{\hat{g}}(\bi{r},\bi{k},\omega_n)
\end{pmatrix},
\end{align}
with the center-of-mass coordinate of a Cooper pair $\bi{r}$,
the direction of the relative momentum of a Cooper pair $\bi{k}$,
and the Matsubara frequency $\omega_n=(2n+1)\pi k_B T$ with $n\in\mathbb{Z}$.
The quasi-classical Green's function satisfies a normalization condition $\widehat{g}^2=-\pi^2\widehat{1}$.
The pair potential is described in particle-hole space by
\begin{align}
\widehat{\Delta }(\bi{r},\bi{k})=
\begin{pmatrix}
0 & \hat{\Delta }(\bi{r},\bi{k}) \\
-\hat{\Delta }^{\dagger }(\bi{r},\bi{k}) & 0
\end{pmatrix},
\end{align}
and that in spin space for the spin-triplet state is defined by the $d$-vector as
\begin{align}
\hat{\Delta }(\bi{r},\bi{k})=(i\hat{\bi{\sigma }}\hat{\sigma }_y)\cdot \bi{d}(\bi{r},\bi{k}),
\end{align}
with the Pauli matrix $\hat{\bi{\sigma }}$.
The Fermi velocity in the Eilenberger equation is given as $\bi{v}(\bi{k})=v_F\bi{k}$ for the 3D Fermi sphere.

We solve eq.~\eqref{Eilenberger eq} by the Riccati method~\cite{schopohl:1995,nagato:1993,eschrig:2000}.
We introduce the Riccati amplitude $\hat{a}=(\hat{1}+\hat{g})^{-1}\hat{f}$ 
and $\hat{b}=(\hat{1}+\underline{\hat{g}})^{-1}\underline{\hat{f}}$
related to particle- and hole-like projections of off-diagonal propagators, respectively.
Equation \eqref{Eilenberger eq} can be rewritten as Riccati equations:
\begin{multline}
\hbar\bi{v}(\bi{k})\cdot\bi{\nabla }\hat{a}(\bi{r},\bi{k},\omega_n) \\
= \hat{\Delta }(\bi{r},\bi{k})-\hat{a}(\bi{r},\bi{k},\omega_n)\hat{\Delta }^{\dagger }(\bi{r},\bi{k}) \hat{a}(\bi{r},\bi{k},\omega_n)\\
\shoveright{-2\omega_n\hat{a}(\bi{r},\bi{k},\omega_n),} \\
\shoveleft{-\hbar\bi{v}(\bi{k})\cdot\bi{\nabla }\hat{b}(\bi{r},\bi{k},\omega_n)} \\
= \hat{\Delta }^{\dagger }(\bi{r},\bi{k})-\hat{b}(\bi{r},\bi{k},\omega_n)\hat{\Delta }(\bi{r},\bi{k})\hat{b}(\bi{r},\bi{k},\omega_n)\\
-2\omega_n\hat{b}(\bi{r},\bi{k},\omega_n).
\label{Riccati eqs}
\end{multline}
These equations are solved by integration toward $\bi{k}$ for $\hat{a}(\bi{r},\bi{k},\omega_n)$
and toward $-\bi{k}$ for $\hat{b}(\bi{r},\bi{k},\omega_n)$.
By the Riccati amplitude, the quasi-classical Green's function is given as
\begin{align}
\widehat{g} = -i\pi
\begin{pmatrix}
(\hat{1}+\hat{a}\hat{b})^{-1} & 0 \\
0 & (\hat{1}+\hat{b}\hat{a})^{-1}
\end{pmatrix}
\begin{pmatrix}
\hat{1}-\hat{a}\hat{b} & 2i\hat{a} \\
-2i\hat{b} & -(\hat{1}-\hat{b}\hat{a})
\end{pmatrix}.
\end{align}

By using the quasi-classical Green's function, the temperature-dependent mass and spin currents are calculated using
\begin{align}
\bi{j}(\bi{r},T) &=
mN_0\pi k_BT\sum_{\omega_n} 
\langle \bi{v}(\bi{k}) \ {\rm Im} \left[ g_0(\bi{r},\bi{k}, \omega_n) \right] \rangle_{\bi{k}}, 
\label{mcurrentT}\\
\bi{j}_s^{\mu }(\bi{r},T) &=
\frac{\hbar }{2}N_0\pi k_BT\sum_{\omega_n} 
\langle \bi{v}(\bi{k}) \ {\rm Im} \left[ g_{\mu }(\bi{r},\bi{k}, \omega_n) \right] \rangle_{\bi{k}},
\label{scurrentT}
\end{align}
respectively, where $N_0$ is the DOS in the normal state, 
$m$ is the mass of the $^3$He atom,
$\langle\cdots\rangle_{\bi{k}}$ indicates the Fermi surface average,
and $g_{\mu }$ is the component of the quasi-classical Green's function $\hat{g}$ in spin space, namely,
\begin{align}
\hat{g} =
\begin{pmatrix}
g_0+g_z & g_x-ig_y \\
g_x+ig_y & g_0-g_z
\end{pmatrix}. \nn
\end{align}
The mass and spin currents and local density of states (LDOS) for the energy $E$ are given by
\begin{align}
\bi{j}(\bi{r},E) &=\left\langle \bi{j}(\bi{r},\bi{k},E)\right\rangle_{\bi{k}} \nn\\
&=\ mN_0\left\langle \bi{v}(\bi{k}) \ {\rm Re} \left[g_0(\bi{r},\bi{k}, \omega_n)|_{i\omega_n \rightarrow E+i\eta}\right] \right\rangle_{\bi{k}},
\label{mcurrentE}\\ 
\bi{j}_s^{\mu }(\bi{r},E) &=\left\langle \bi{j}_s^{\mu }(\bi{r},\bi{k},E)\right\rangle_{\bi{k}} \nn\\
&=\ \frac{\hbar }{2}N_0\left\langle \bi{v}(\bi{k}) \ {\rm Re} \left[g_{\mu }(\bi{r},\bi{k}, \omega_n)|_{i\omega_n \rightarrow E+i\eta}\right] \right\rangle_{\bi{k}},
\label{scurrentE}\\ 
N(\bi{r},E) &= \left\langle N(\bi{r},\bi{k},E)\right\rangle_{\bi{k}} \nn\\
&=\ N_0 \left\langle {\rm Re} \left[g_0(\bi{r},\bi{k}, \omega_n)|_{i\omega_n \rightarrow E+i\eta}\right] \right\rangle_{\bi{k}},
\label{LDOS}
\end{align}
respectively, where $\eta$ is a positive infinitesimally small constant.

\section{A-Phase}
\label{sec:A}

In this section, we consider the superfluid $^3$He A-phase in a slab 
with a small thickness along the $z$-direction.
In the sufficiently thin slab, the $l$-vector pointing to the direction of the angular momentum of a Cooper pair is aligned toward the $z$-direction and
the $d$-vector is also aligned toward the $z$-direction by the dipole interaction~\cite{vollhardt:book}.
This system is realized using a slab of sub-$\mu$m thickness~\cite{tsutsumi:2011b},
which is already realized experimentally~\cite{bennett:2010}.
The uniform pair potential in the system is described by the $d$-vector as
\begin{align}
\bi{d}(\bi{k})=\Delta_A(k_x+ik_y)\bi{z},
\label{chiral}
\end{align}
where $\bi{z}$ is a unit vector and $\Delta_A$ is the amplitude of the superfluid gap in the A-phase.

Here, we discuss a side edge of the slab at $x=0$ which is filled with the superfluid $^3$He at $x>0$.
Assuming that the side edge is specular,
only the $k_x$-component of the $d$-vector is suppressed at the edge 
and the $d$-vector must recover to the chiral state in the bulk as eq.~\eqref{chiral} far from the edge.
Accordingly, we solve the Riccati equations in eq.~\eqref{Riccati eqs} under the pair potential with
\begin{align}
d_z(x,\bi{k})=\Delta_A\left[k_x\tanh\left(\frac{x}{\xi_A}\right)+ik_y\right],
\label{OPA}
\end{align}
where the coherent length is defined by $\xi_A\equiv\hbar v_F/\Delta_A$.
This $d$-vector form embodies the fact that the $k_x$- and $k_y$-components are suppressed and intact at the edge, respectively.
Since the pair potential in spin space is
\begin{align*}
\hat{\Delta }=
\begin{pmatrix}
0 & d_z \\
d_z & 0
\end{pmatrix},
\end{align*}
we can obtain the Riccati amplitude in spin space as
\begin{align*}
\hat{a}=
\begin{pmatrix}
0 & a \\
a & 0
\end{pmatrix},\
\hat{b}=
\begin{pmatrix}
0 & b \\
b & 0
\end{pmatrix}.
\end{align*}
Thus, we can reduce the matrix Riccati equations in eq.~\eqref{Riccati eqs} to the scalar Riccati equations
\begin{align}
\hbar v_Fk_x\frac{\partial }{\partial x}a=&d_z-d_z^*a^2-2\omega_na, \nn\\
-\hbar v_Fk_x\frac{\partial }{\partial x}b=&d_z^*-d_zb^2-2\omega_nb.
\end{align}
The solution of the Riccati amplitude is
\begin{align}
a(x,\bi{k},\omega_n)=&\quad\ \frac{\omega_n-\sqrt{\omega_n^2+\Delta_A^2\sin^2\theta }+d_z(x,\bi{k})}
{\omega_n+\sqrt{\omega_n^2+\Delta_A^2\sin^2\theta }-d_z^*(x,\bi{k})},\nn\\
b(x,\bi{k},\omega_n)=&-\frac{\omega_n-\sqrt{\omega_n^2+\Delta_A^2\sin^2\theta }-d_z^*(x,\bi{k})}
{\omega_n+\sqrt{\omega_n^2+\Delta_A^2\sin^2\theta }+d_z(x,\bi{k})},
\label{RiccatiA}
\end{align}
where we use $k_x=\sin\theta\cos\phi$ and $k_y=\sin\theta\sin\phi$
in the spherical coordinates on the unit Fermi surface.
This solution formally satisfies the boundary condition~\cite{schopohl:cond}, namely,
\begin{align}
a(-\infty,\bi{k},\omega_n)=&\frac{d_z(-\infty,\bi{k})}{\omega_n+\sqrt{\omega_n^2+|d_z(-\infty,\bi{k})|^2}},\nn\\
b(+\infty,\bi{k},\omega_n)=&\frac{d_z^*(+\infty,\bi{k})}{\omega_n+\sqrt{\omega_n^2+|d_z(+\infty,\bi{k})|^2}}.
\end{align}
Moreover, this is a self-consistent solution in the weak-coupling limit as demonstrated in Appendix.
The quasi-classical Green's function is obtained using the relation $\hat{g}=(\hat{1}+\hat{a}\hat{b})^{-1}(\hat{1}-\hat{a}\hat{b})$.
Finally, the spin component of quasi-classical Green's function $g_0=(1-ab)/(1+ab)$ is only finite in $\hat{g}$ as
\begin{multline}
g_0(x,\bi{k},\omega_n)=\frac{1}{\sqrt{\omega_n^2+\Delta_A^2\sin^2\theta }}\\
\times\left[\omega_n+\frac{\Delta_A^2\sin^2\theta\cos^2\phi }
{2(\omega_n+i\Delta_A\sin\theta\sin\phi)}\ {\rm sech}^2\left(\frac{x}{\xi_A}\right)\right].
\label{Green}
\end{multline}

\subsection{Local density of states}

From the quasi-classical Green's function in eq.~\eqref{Green}, 
we can calculate $\theta$-angle-resolved LDOS using eq.~\eqref{LDOS} as
\begin{align}
N(x,\theta,E)\equiv\int\frac{d\phi }{2\pi }N(x,\bi{k},E)
=\frac{N_0}{2}\ {\rm sech}^2\left(\frac{x}{\xi_A }\right),
\label{LDOSMJA}
\end{align}
for the Majorana bound state $|E|<\Delta_A\sin\theta$ and
\begin{multline}
N(x,\theta,E)=N_0\left[\frac{|E|}{\sqrt{E^2-\Delta_A^2\sin^2\theta }}\right.\\
\left.-\frac{1}{2}\left(\frac{|E|}{\sqrt{E^2-\Delta_A^2\sin^2\theta }}-1\right)
\ {\rm sech}^2\left(\frac{x}{\xi_A }\right)\right],
\label{LDOScontA}
\end{multline}
for the continuum state $|E|>\Delta_A\sin\theta$.
Thus, QPs feel the pair potential $\Delta_A\sin\theta$
where $\theta$ is the polar angle from the point node situated at the pole of the Fermi sphere.
The $\theta$-angle-resolved LDOS at the edge $x=0$ for $\theta=\pi/2$ is shown in Fig.~\ref{energyA}(a).
The LDOS from the Majorana zero energy mode has a constant and finite value, $N_0/2$.
A similar value is also obtained by numerical calculation in a finite temperature~\cite{tsutsumi:2010b,tsutsumi:2011b}.
The $\theta$-angle-resolved LDOS from Majorana QPs is independent of the polar angle $\theta$.
This indicates that the dispersion of Majorana QPs forms a ``Majorana valley"~\cite{tsutsumi:2010b,tsutsumi:2011b}.
The LDOS in the continuum state is also deformed to conserve the DOS against the appearance of Majorana QPs.

\subsection{Edge mass current and angular momentum}

\begin{figure}
\begin{center}
\includegraphics[width=5cm]{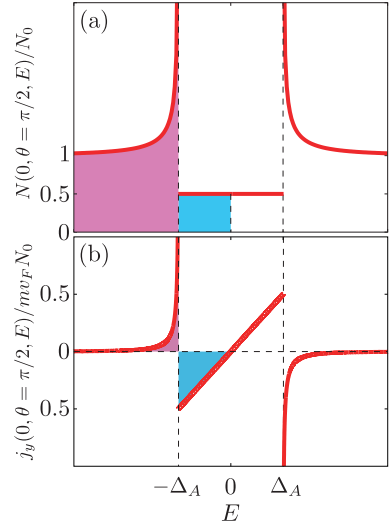}
\end{center}
\caption{\label{energyA}(Color online) 
Energy profiles of $\theta$-angle-resolved LDOS (a) and mass current along the edge (b) at $x=0$ for $\theta=\pi/2$ in the A-phase.
At a zero temperature, QPs fill the colored (shaded) states in (a).
The mass currents from the bound and continuum states are derived 
by integrating the blue (light gray) and pink (gray) regions in (b), respectively.
}
\end{figure}

We can also calculate $\theta$-angle-resolved mass current along the edge using eq.~\eqref{mcurrentE} as
\begin{align}
j_y(x,\theta,E)&\equiv\int\frac{d\phi }{2\pi }j_y(x,\bi{k},E)\nn\\
&=\frac{mv_FN_0}{2}\frac{E}{\Delta_A}\ {\rm sech}^2\left(\frac{x}{\xi_A }\right),
\label{jEMJ}
\end{align}
for the bound state $|E|<\Delta_A\sin\theta$ and as
\begin{multline}
j_y(x,\theta,E)=-\frac{mv_FN_0}{4}\frac{E}{|E|}\left[\frac{\sqrt{E^2-\Delta_A^2\sin^2\theta }}{\Delta_A}\right.\\
\left.+\frac{E^2}{\Delta_A\sqrt{E^2-\Delta_A^2\sin^2\theta }}-2\frac{|E|}{\Delta_A}\right]\ {\rm sech}^2\left(\frac{x}{\xi_A }\right),
\label{jE}
\end{multline}
for the continuum state $|E|>\Delta_A\sin\theta$.
The $\theta$-angle-resolved mass current at the edge for $\theta=\pi/2$ is shown in Fig.~\ref{energyA}(b).
The energy spectrum of the mass current from Majorana QPs is linear.
The edge mass current from the continuum state is due to the deformation of the LDOS,
which decreases away from $\Delta_A\sin\theta$.
The asymptotic behavior of the edge mass current in eq.~\eqref{jE} at $x=0$ is estimated as
\begin{align}
j_y(x=0,\theta, E)\approx -\frac{mv_FN_0}{16}\sin\theta\left(\frac{\Delta_A\sin\theta}{E}\right)^3
\label{japprox}
\end{align}
for $|E|\gg\Delta_A\sin\theta$, implying that the contribution of the QPs to the edge mass current decreases as $\sim E^{-3}$.
This power functional behavior of the decrease means that 
QPs contributing to the edge mass current are not confined in only the vicinity of the Fermi surface.

Since QPs fill the energy state up to the Fermi energy at a zero temperature,
the mass current along the edge from the Majorana bound state is obtained as
\begin{align}
j_y^{\rm MJ}(x)&=\left\langle\int_{-\Delta_A\sin\theta }^{0}dEj_y(x,\bi{k},E)\right\rangle_{\bi{k}}\nn\\
&=-\frac{mv_FN_0\Delta_A}{6}\ {\rm sech}^2\left(\frac{x}{\xi_A }\right),
\label{jMJ}
\end{align}
and that from the continuum state is obtained as
\begin{align}
j_y^{\rm cont}(x)&=\left\langle\int_{-\infty }^{-\Delta_A\sin\theta }dEj_y(x,\bi{k},E)\right\rangle_{\bi{k}}\nn\\
&=\frac{mv_FN_0\Delta_A}{12}\ {\rm sech}^2\left(\frac{x}{\xi_A }\right).
\end{align}
The total mass current along the edge from the Majorana bound state is
\begin{align}
J_y^{\rm MJ}\equiv\int_0^{\infty }dxj_y^{\rm MJ}(x)=-\frac{n\hbar }{2},
\label{JMJ}
\end{align}
and that from the continuum state is
\begin{align}
J_y^{\rm cont}\equiv\int_0^{\infty }dxj_y^{\rm cont}(x)=\frac{n\hbar }{4},
\label{Jcont}
\end{align}
where the density of $^3$He atoms $n$ emerges from the normal DOS $N_0=(3/mv_F^2)n$.
The same result with eq.~\eqref{JMJ} is obtained in connection with the chiral superconductor Sr$_2$RuO$_4$~\cite{furusaki:2001}.
Since these currents flow oppositely, the total mass current induced by the edge state is
\begin{align}
J_y=J_y^{\rm MJ}+J_y^{\rm cont}=-\frac{n\hbar }{4}.
\label{J}
\end{align}

In a disk with a large radius $R\gg\xi$,
since the mass current can be regarded as localized at the edge,
the angular momentum from each state is calculated as 
\begin{align}
L_z^{\rm MJ}=N\hbar,\ L_z^{\rm cont}=-\frac{N\hbar }{2}.
\label{Lzsep}
\end{align}
Finally, the total angular momentum simply becomes
\begin{align}
L_z=L_z^{\rm MJ}+L_z^{\rm cont}=\frac{N\hbar }{2}.
\label{Lz}
\end{align}
The cancelled angular momentum $N\hbar/2$ may be related to our finding
that the DOS of Majorana QPs is $N_0/2$.
The total angular momentum due to the edge mass current coincides with that in the 2D A-phase under a uniform pair potential~\cite{stone:2004,sauls:2011}.
Interestingly, half of the angular momentum from the Majorana bound state is canceled by that from the continuum state.

\subsection{Temperature dependence}

\begin{figure}
\begin{center}
\includegraphics[width=6.5cm]{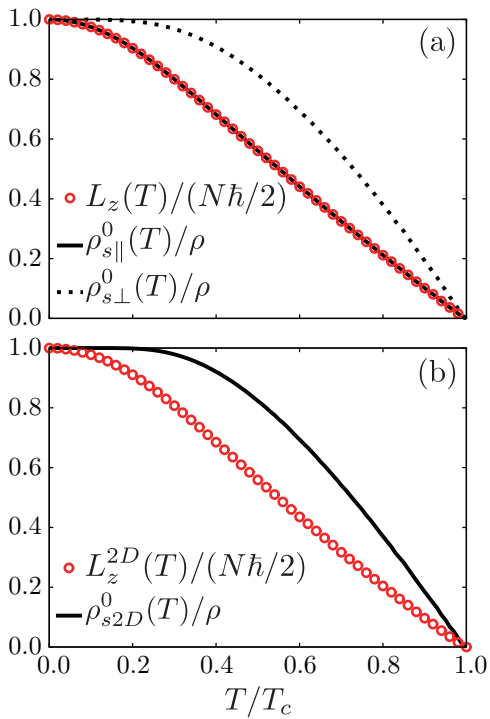}
\end{center}
\caption{\label{temperatureA}(Color online)
Temperature dependence of angular momentum (open circles) 
in the A-phase (a) and 2D chiral $p$-wave state (b) with the superfluid density.
The components of the superfluid density tensor $\rho_{s\parallel }^0$ (solid line) and $\rho_{s\perp }^0$ (dotted line) are shown in (a) 
and the superfluid density $\rho_{s2D}^0$ (solid line) is shown in (b).
$L_z(T)$ and $\rho_{s\parallel }^0(T)$ in (a) completely coincide.
}
\end{figure}

The temperature dependence of the angular momentum $L_z(T)$ by the edge mass current 
is calculated using eq.~\eqref{mcurrentT} as
\begin{multline}
L_z(T)=\frac{3}{4}N\hbar\frac{\pi k_BT}{\Delta_A}\\
\times\sum_{\omega_n}
\left[\frac{3\omega_n^2+\Delta_A^2}{\Delta_A^2}\sin^{-1}\frac{\Delta_A}{\sqrt{\omega_n^2+\Delta_A^2}}
-3\frac{|\omega_n|}{\Delta_A}\right].
\label{LzT}
\end{multline}
This temperature dependence is shown in Fig.~\ref{temperatureA}(a) by open circles.
The component of the superfluid density tensor 
parallel (perpendicular) to the direction of the point nodes $\rho_{s\parallel }^0$ ($\rho_{s\perp }^0$)~\cite{cross:1975} is also depicted by two lines.
The angular momentum $L_z(T)$ (open circles) has the same temperature dependence of $\rho_{s\parallel }^0(T)$ (solid line), as pointed out by Kita~\cite{kita:1998}.
This complete correspondence of the temperature dependence, however, may be accidental 
because the low-temperature depletion of the angular momentum can be explained by only the Majorana QPs, as mentioned below.
Moreover, in the 2D Fermi surface model, the temperature dependence of the angular momentum has no connection with that of the superfluid density [see Fig.~\ref{temperatureA}(b)].

According to eq.~\eqref{LzT}, we can derive the low-temperature behavior of the angular momentum as
\begin{align}
L_z(T)=\frac{N\hbar }{2}\left[1-\left(\frac{\pi k_BT}{\Delta_A}\right)^2+\mathrm{O}\left(\frac{\pi k_BT}{\Delta_A}\right)^4\right].
\label{LzTlow}
\end{align}
The low-temperature depletion of the angular momentum contains both contributions
of the excitations of Majorana QPs and point nodes.
Majorana QPs contribute to the depletion as that in $T^2$-power
because $j_y(x,\theta,E)$ in the Majorana bound state has a linear energy dependence [Fig.~\ref{energyA}(b)].
The contribution of the excitations at point nodes is derived from 
the low-energy behavior of the energy spectrum of the mass current in the continuum state.
The energy spectrum at $x=0$ is obtained using eq.~\eqref{jE} as
\begin{multline}
j_y(x=0,E)=\int_0^{\theta_0}\sin\theta d\theta j_y(x=0,\theta,E)\\
=-\frac{mv_FN_0}{16}\left[\left\{3\left(\frac{E}{\Delta_A}\right)^2-1\right\}\ln\frac{1+(E/\Delta_A)}{1-(E/\Delta_A)}\right.\\
\left.-6\frac{E}{\Delta_A}+8\frac{E}{\Delta_A}\sqrt{1-\left(\frac{E}{\Delta_A}\right)^2}\right],
\end{multline}
for $|E|<\Delta_A$, where $\theta_0=\sin^{-1}(|E|/\Delta_A)$.
This low-energy behavior is estimated as
\begin{align}
j_y(x=0,E)\approx -\frac{mv_FN_0}{12}\left(\frac{E}{\Delta_A}\right)^3,
\end{align}
for $|E|\ll\Delta_A$.
Thus, the excitations at point nodes contribute to the angular momentum as the fourth order of temperature ($\sim T^4$).
Thus, the observation of the depletion of the angular momentum as that in $T^2$-power could establish the existence of Majorana QPs.

In the 2D Fermi surface model, where point nodes are absent,
LDOS $N(x,E)$ is obtained by the substitution of $\Delta_{2D}$ for $\Delta_A\sin\theta$ on the right-hand side of eqs.~\eqref{LDOSMJA} and \eqref{LDOScontA},
where $\Delta_{2D}$ is the amplitude of the superfluid gap in the 2D chiral $p$-wave state.
The values of the total mass current and angular momentum are also given by eqs.~\eqref{JMJ}-\eqref{J} and eqs.~\eqref{Lzsep} and \eqref{Lz}, respectively.
However, the temperature dependence of the angular momentum is different from eq.~\eqref{LzT}, that is, it is given by
\begin{multline}
L_z^{2D}(T)=N\hbar\frac{\pi k_BT}{\Delta_{2D}}\\
\times\sum_{\omega_n}
\left[\frac{\sqrt{\omega_n^2+\Delta_{2D}^2}}{\Delta_{2D}} + \frac{\omega_n^2}{\Delta_{2D}\sqrt{\omega_n^2+\Delta_{2D}^2}}
-2\frac{|\omega_n|}{\Delta_{2D}}\right].
\end{multline}
This temperature dependence is also shown in Fig.~\ref{temperatureA}(b) with the superfluid density in the 2D chiral $p$-wave state $\rho_{s2D}^0$.
These behaviors are clearly different as pointed out by Sauls~\cite{sauls:2011}
because the low-temperature depletion of the angular momentum is due to the excitations of Majorana QPs.
The low-temperature behavior of the angular momentum is
\begin{align}
L_z^{2D}(T)=\frac{N\hbar }{2}\left[1-\frac{2}{3}\left(\frac{\pi k_BT}{\Delta_{2D}}\right)^2+\mathrm{O}\left(\frac{\pi k_BT}{\Delta_{2D}}\right)^4\right].
\label{Lz2DTlow}
\end{align}
The energy spectrum of the mass current $j_y(x,E)$ in the Majorana bound state is obtained
by the substitution of $\Delta_{2D}$ for $\Delta_A$ on the right-hand side of eq.~\eqref{jEMJ}.
The difference between the coefficients on the second order of temperature in eqs.~\eqref{LzTlow} and \eqref{Lz2DTlow} comes from the difference in the normal DOS,
where $N_0=(2/mv_F^2)n$ for the 2D Fermi surface is $2/3$ of that for the 3D Fermi sphere.
Although the coefficients are different, the temperature dependences of the angular momentum are similar in these models owing to $\Delta_A>\Delta_{2D}$.

\subsection{Intrinsic angular momentum}

The angular momentum by the edge mass current at a zero temperature is $L=N\hbar/2$.
This magnitude corresponds to one of the predictions of the intrinsic angular momentum~\cite{ishikawa:1977,ishikawa:1980,volovik:1995}.
The magnitude of the intrinsic angular momentum in this prediction is expected naively
when all Cooper pairs carry one unit of angular momentum.
However, the edge mass current is carried by a portion of QPs
because the contribution of QPs to the edge mass current decreases as $E^{-3}$ away from the superfluid gap.
QPs carrying the edge mass current are distributed
neither in the narrow energy shell around the Fermi level nor up to the bottom of the Fermi sea.
The total number of $^3$He atoms $N$ in the angular momentum by the edge mass current comes from the normal DOS $N_0=(3/mv_F^2)n$.
Moreover, the temperature dependence of the angular momentum by the edge mass current has no connection with that of the superfluid density, which is related to the number of Cooper pairs.
The low-temperature depletion of this angular momentum as that in $T^2$-power is due to the excitations of Majorana QPs in the edge state.

\section{B-Phase}
\label{sec:B}

The uniform pair potential in the superfluid $^3$He B-phase is described by the $d$-vector as
\begin{align}
\bi{d}(\bi{k})=\Delta_BR(\bi{n},\theta_d)(k_x\bi{x}+k_y\bi{y}+k_z\bi{z}),
\label{dB}
\end{align}
where $\bi{x}$, $\bi{y}$, and $\bi{z}$ are unit vectors, 
$\Delta_B$ is the amplitude of the superfluid gap in the B-phase,
and $R(\bi{n},\theta_d)$ is a rotation matrix with a rotation axis $\bi{n}$ and a rotation angle $\theta_d$ about $\bi{n}$.
The rotation matrix gives the relative angle between $\bi{k}$ and $\bi{d}$.
The spin state is stable by the dipole interaction when $\theta_d=\theta_L\equiv\cos^{-1}(-1/4)$ and the $n$-vector is perpendicular to a surface~\cite{vollhardt:book}.

In this section, we consider the superfluid $^3$He B-phase filled in $z>0$ with a specular edge at $z=0$.
Only the $k_z$-component of the $d$-vector is suppressed near the edge within the coherent length 
and the $d$-vector must recover to the bulk form in eq.~\eqref{dB} far from the edge.
In this system, the $n$-vector points to the $z$-direction perpendicular to the edge,
and the angle $\theta_d$ remains intact at the bulk value $\theta_L$ because the angle can vary not on the order of the coherent length but over a larger length, namely, the dipole coherent length~\cite{vollhardt:book}.
Accordingly, the pair potential with a specular edge at $z=0$ is 
\begin{align}
\hat{\Delta }(z,\bi{k})=\Delta_B
\begin{pmatrix}
-\sin\theta e^{-i(\phi+\theta_L)} & \cos\theta\tanh(z/\xi_B) \\
\cos\theta\tanh(z/\xi_B) & \sin\theta e^{i(\phi+\theta_L)}
\end{pmatrix},
\label{OPB}
\end{align}
where we use $k_x=\sin\theta\cos\phi$, $k_y=\sin\theta\sin\phi$, and $k_z=\cos\theta$
and the coherent length is defined by $\xi_B\equiv\hbar v_F/\Delta_B$.

The Eilenberger equation eq.~\eqref{Eilenberger eq} can be solved similarly to the A-phase after the unitary transformation using the unitary matrix~\cite{mizushima:private}
\begin{align}
\widehat{M}=
\begin{pmatrix}
\hat{M} & 0 \\
0 & \hat{M}^*
\end{pmatrix},\
\hat{M}=\frac{1}{\sqrt{2}}
\begin{pmatrix}
u & u^* \\
u & -u^*
\end{pmatrix},
\end{align}
where $u^2=-ie^{i(\phi+\theta_L)}$.
The unitary-transformed pair potential is 
\begin{align}
\widehat{\Delta }'\equiv\widehat{M}\widehat{\Delta }\widehat{M}^{\dagger }=
\begin{pmatrix}
0 & \hat{\Delta }' \\
-\hat{\Delta }'^{\dagger } & 0
\end{pmatrix},
\end{align}
where
\begin{align}
\hat{\Delta }'\equiv\hat{M}\hat{\Delta }\hat{M}^T=
\begin{pmatrix}
d'(z,\theta) & 0 \\
0 & -d'^*(z,\theta)
\end{pmatrix},
\label{utOPBsp}
\end{align}
with
\begin{align}
d'(z,\theta)=\Delta_B\left[\cos\theta\tanh\left(\frac{z}{\xi_B}\right)+i\sin\theta\right].
\end{align}
Finally, the unitary-transformed Eilenberger equation is
\begin{multline}
-i\hbar v_F\cos\theta\frac{\partial }{\partial z}\widehat{g}'(z,\theta,\omega_n)\\
=\left[i\omega_n\widehat{\sigma }_z-\widehat{\Delta }'(z,\theta),
\widehat{g}'(z,\theta,\omega_n) \right],
\end{multline}
with $\widehat{g}'\equiv\widehat{M}\widehat{g}\widehat{M}^{\dagger }$,
where $\widehat{M}$ and $\widehat{\sigma }_z$ are commutable.

Since the unitary-transformed pair potential in spin space is described by eq.~\eqref{utOPBsp},
we can obtain the Riccati amplitude in spin space:
\begin{align*}
\hat{a}'=
\begin{pmatrix}
a' & 0 \\
0 & -a'^*
\end{pmatrix},\
\hat{b}'=
\begin{pmatrix}
b' & 0 \\
0 & -b'^*
\end{pmatrix}.
\end{align*}
Owing to the analogy of the A-phase in \S\ref{sec:A}, the solution of the Riccati amplitude is obtained by the substitution of $\Delta_B$ and $d'$ for $\Delta_A\sin\theta$ and $d_z$, respectively, as
\begin{align}
a'(z,\theta,\omega_n)=&\quad\ \frac{\omega_n-\sqrt{\omega_n^2+\Delta_B^2}+d'(z,\theta)}
{\omega_n+\sqrt{\omega_n^2+\Delta_B^2}-d'^*(z,\theta)},\nn\\
b'(z,\theta,\omega_n)=&-\frac{\omega_n-\sqrt{\omega_n^2+\Delta_B^2}-d'^*(z,\theta)}
{\omega_n+\sqrt{\omega_n^2+\Delta_B^2}+d'(z,\theta)}.
\label{RiccatiB}
\end{align}
The quasi-classical Green's function is obtained by the unitary transformation of this solution as $\widehat{g}=\widehat{M}^{\dagger }\widehat{g}'\widehat{M}$.
The obtained quasi-classical Green's function is self-consistent in the weak-coupling limit similarly to the A-phase case, as discussed in Appendix.
The spin components of quasi-classical Green's function $\hat{g}=\hat{M}^{\dagger }\hat{g}'\hat{M}$ are
\begin{align}
&g_0={\rm Re}\left[\frac{1-a'b'}{1+a'b'}\right],\ g_x=i\sin(\phi+\theta_L)\ {\rm Im}\left[\frac{1-a'b'}{1+a'b'}\right],\nn\\
&g_y=-i\cos(\phi+\theta_L)\ {\rm Im}\left[\frac{1-a'b'}{1+a'b'}\right],\ g_z=0.
\end{align}
Since
\begin{multline}
\frac{1-a'b'}{1+a'b'}=\frac{1}{\sqrt{\omega_n^2+\Delta_B^2}}\\
\times\left[\omega_n+\frac{\Delta_B^2\cos^2\theta }
{2(\omega_n+i\Delta_B\sin\theta)}\ {\rm sech}^2\left(\frac{z}{\xi_B}\right)\right],
\end{multline}
by the analogy of the A-phase in eq.~\eqref{Green}, we obtain
\begin{widetext}
\begin{align}
g_0(z,\bi{k},\omega_n)=&\frac{1}{\sqrt{\omega_n^2\!+\!\Delta_B^2}}
\!\left[\omega_n\!+\!\frac{\Delta_B^2\cos^2\theta }{4}
\!\left(\frac{1}{\omega_n\!+\!i\Delta_B\sin\theta }\!+\!\frac{1}{\omega_n\!-\!i\Delta_B\sin\theta }\right)\!{\rm sech}^2\!\left(\frac{z}{\xi_B}\right)\!\right]\!,
\label{Green0B}\\
g_x(z,\bi{k},\omega_n)=&\quad\ \frac{\sin(\phi\!+\!\theta_L)}{\sqrt{\omega_n^2\!+\!\Delta_B^2}}\frac{\Delta_B^2\cos^2\theta }{4}
\left(\frac{1}{\omega_n\!+\!i\Delta_B\sin\theta }\!-\!\frac{1}{\omega_n\!-\!i\Delta_B\sin\theta }\right)\ {\rm sech}^2\left(\frac{z}{\xi_B}\right),\\
g_y(z,\bi{k},\omega_n)=&-\frac{\cos(\phi\!+\!\theta_L)}{\sqrt{\omega_n^2\!+\!\Delta_B^2}}\frac{\Delta_B^2\cos^2\theta }{4}
\left(\frac{1}{\omega_n\!+\!i\Delta_B\sin\theta }\!-\!\frac{1}{\omega_n\!-\!i\Delta_B\sin\theta }\right)\ {\rm sech}^2\left(\frac{z}{\xi_B}\right),\\
g_z(z,\bi{k},\omega_n)=&\ 0.
\label{GreenzB}
\end{align}

\subsection{Local density of states}

From the quasi-classical Green's function in eq.~\eqref{Green0B}, 
we can calculate LDOS by eq.~\eqref{LDOS} as
\begin{align}
N(z,E)=\frac{\pi }{4}N_0\frac{|E|}{\Delta_B}\ {\rm sech}^2\left(\frac{z}{\xi_B}\right),
\label{LDOSMJB}
\end{align}
for the Majorana bound state $|E|<\Delta_B$ and as
\begin{align}
N(z,E)\!=\!N_0\left[\frac{|E|}{\sqrt{E^2\!-\!\Delta_B^2}}
\!-\!\frac{1}{2}\left(\frac{|E|}{\sqrt{E^2\!-\!\Delta_B^2}}\!-\!\frac{|E|}{\Delta_B}\tan^{-1}\frac{\Delta_B}{\sqrt{E^2\!-\!\Delta_B^2}}\right)
\ {\rm sech}^2\left(\frac{z}{\xi_B}\right)\right],
\label{LDOScontB}
\end{align}
\end{widetext}

for the continuum state $|E|>\Delta_B$.
The LDOS at the edge $z=0$ is shown in Fig.~\ref{energyB}(a).
This LDOS in the Majorana bound state has a linear energy dependence with a slope $(\pi/4)N_0$.
This linear dependence is also obtained by numerical calculation~\cite{buchholtz:1981,nagato:1998,tsutsumi:2011b}.
A ``Majorana cone" formed by the dispersion of Majorana QPs~\cite{chung:2009,nagato:2009} is the cause of this linear dependence.
The LDOS in the continuum state is also deformed to conserve the DOS against the appearance of Majorana QPs.

\subsection{Edge spin current}

\begin{figure}
\begin{center}
\includegraphics[width=5cm]{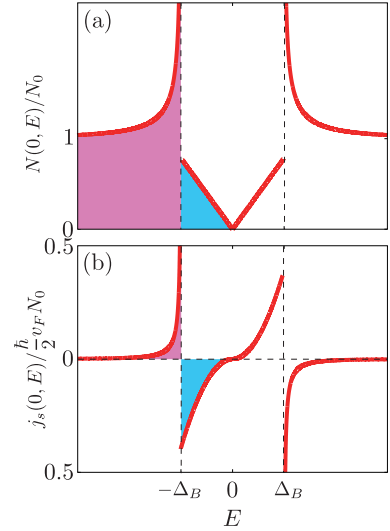}
\end{center}
\caption{\label{energyB}(Color online) 
Energy profiles of LDOS (a) and spin current function (b) at $z=0$ in the B-phase.
At a zero temperature, QPs fill the colored (shaded) states in (a).
The spin currents from the bound and continuum states are derived 
by integrating the blue (light gray) and pink (gray) regions in (b), respectively.
}
\end{figure}

The edge mass current is absent in the superfluid $^3$He B-phase.
This is clearly seen from eq.~\eqref{mcurrentT} because $g_0$ is a real function.
We can calculate the spin current using eq.~\eqref{scurrentE}.
The finite spin currents along the edge are
\begin{align}
\bi{j}_s^x(z,E)=&\quad\ j_s(z,E)(\sin\theta_L\bi{x}+\cos\theta_L\bi{y}),\nn\\
\bi{j}_s^y(z,E)=&-j_s(z,E)(\cos\theta_L\bi{x}-\sin\theta_L\bi{y}),
\end{align}
where $\bi{x}$ and $\bi{y}$ are unit vectors.
The spin current function $j_s(z,E)$ is
\begin{align}
j_s(z,E)=\frac{\pi }{8}\frac{\hbar }{2}v_FN_0\frac{E|E|}{\Delta_B^2}\ {\rm sech}^2\left(\frac{z}{\xi_B}\right),
\end{align}
\begin{widetext}
for the bound state $|E|<\Delta_B$ and
\begin{align}
j_s(z,E)=-\frac{1}{12}\frac{\hbar }{2}v_FN_0\frac{E}{|E|}\left[\frac{\sqrt{E^2-\Delta_B^2}}{\Delta_B}+2\frac{E^2}{\Delta_B\sqrt{E^2-\Delta_B^2}}
-3\frac{E^2}{\Delta_B^2}\tan^{-1}\frac{\Delta_B}{\sqrt{E^2-\Delta_B^2}}\right]\ {\rm sech}^2\left(\frac{z}{\xi_B}\right),
\label{jscont}
\end{align}
\end{widetext}
for the continuum state $|E|>\Delta_B$.
This spin current function at the edge is shown in Fig.~\ref{energyB}(b).
The energy spectrum of the spin current from Majorana QPs is quadratic.
The edge spin current from the continuum state is due to the deformation of the LDOS, which decreases away from $\Delta_B$.
The asymptotic behavior of the spin current function in eq.~\eqref{jscont} at $z=0$ is estimated as
\begin{align}
j_s(z=0,E)\approx -\frac{1}{30}\frac{\hbar }{2}v_FN_0\left(\frac{\Delta_B}{E}\right)^3,
\end{align}
for $|E|\gg\Delta_B$, implying that the contribution of QPs decreases with the same power law $\sim E^{-3}$, as in the A-phase.
This power law behavior of such a decrease means that
QPs carrying the edge spin current are distributed neither in the narrow energy shell around the Fermi level nor up to the bottom of the Fermi sea.

Since QPs fill the energy state up to the Fermi energy at a zero temperature,
the spin current from the Majorana bound state is obtained as
\begin{align}
j_s^{\rm MJ}(z)&=\int_{-\Delta_B}^{0}dEj_s(z,E)\nn\\
&=-\frac{\pi }{24}\frac{\hbar }{2}v_FN_0\Delta_B\ {\rm sech}^2\left(\frac{z}{\xi_B}\right),
\label{jsMJ}
\end{align}
and that from the continuum state is 
\begin{align}
j_s^{\rm cont}(z)&=\int_{-\infty }^{-\Delta_B}dEj_s(z,E)\nn\\
&=\frac{\pi }{24}\left(1-\frac{4}{3\pi }\right)\frac{\hbar }{2}v_FN_0\Delta_B\ {\rm sech}^2\left(\frac{z}{\xi_B}\right).
\end{align}
The total spin current from the Majorana bound state is
\begin{align}
J_s^{\rm MJ}\equiv\int_0^{\infty }dzj_s^{\rm MJ}(z)
=-\frac{\pi }{8}\frac{\kappa }{2\pi }n\hbar,
\label{JsMJ}
\end{align}
and that from the continuum state is
\begin{align}
J_s^{\rm cont}\equiv\int_0^{\infty }dzj_s^{\rm cont}(z)
=\frac{\pi }{8}\left(1-\frac{4}{3\pi }\right)\frac{\kappa }{2\pi }n\hbar,
\label{Jscont}
\end{align}
where the quantum of circulation $\kappa=h/(2m)$ emerges from the ratio of coefficients between the spin and mass currents.
Since the edge currents from the Majorana bound state and continuum state flow oppositely, similarly to the A-phase, the total spin current by the edge state is
\begin{align}
J_s=J_s^{\rm MJ}+J_s^{\rm cont}=-\frac{\kappa }{2\pi }\frac{n\hbar }{6}.
\label{Js}
\end{align}
The numerical coefficient of $n\hbar$ in the total spin current is $2/3$ of that in the total mass current 
because QPs contributing to the spin current are QPs in two of three spin states.
Finally, each spin component of the total spin current is
\begin{align}
\bi{J}_s^x=&-\frac{\kappa }{2\pi }\frac{n\hbar }{6}(\sin\theta_L\bi{x}+\cos\theta_L\bi{y}),\nn\\
\bi{J}_s^y=&\quad\ \frac{\kappa }{2\pi }\frac{n\hbar }{6}(\cos\theta_L\bi{x}-\sin\theta_L\bi{y}).
\end{align}

\subsection{Temperature dependence}

\begin{figure}
\begin{center}
\includegraphics[width=6.5cm]{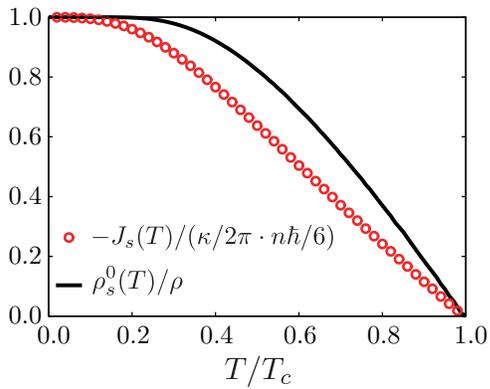}
\end{center}
\caption{\label{temperatureB}(Color online)
Temperature dependence of total spin current (open circles) in the B-phase with the superfluid density (solid line).
}
\end{figure}

The temperature dependence of the total spin current $J_s(T)$ is calculated using eq.~\eqref{scurrentT} as
\begin{multline}
J_s(T)=-\frac{\kappa }{2\pi }\frac{n\hbar }{4}\frac{\pi k_BT}{\Delta_B}
\sum_{\omega_n}\left[\frac{\sqrt{\omega_n^2+\Delta_B^2}}{\Delta_B}\right. \\
\left.+2\frac{\omega_n^2}{\Delta_B\sqrt{\omega_n^2+\Delta_B^2}}
+3\frac{\omega_n^2}{\Delta_B^2}\ln\frac{\sqrt{\omega_n^2+\Delta_B^2}-\Delta_B}{|\omega_n|}\right].
\label{JsT}
\end{multline}
This temperature dependence is shown in Fig.~\ref{temperatureB} by the open circles
compared with the superfluid density $\rho_s^0$ by the line.
These behaviors are clearly different because the low-temperature depletion of the total spin current is due to the excitations of Majorana QPs.
There is no other low energy excitation.

According to eq.~\eqref{JsT}, 
we can derive the low-temperature behavior of the total spin current as
\begin{align}
J_s(T)=-\frac{\kappa }{2\pi }\frac{n\hbar }{6}\left[1-C\left(\frac{\pi k_BT}{\Delta_B}\right)^3+\mathrm{O}\left(\frac{\pi k_BT}{\Delta_B}\right)^4\right],
\end{align}
where the coefficient $C$ is fixed within $3/5\le C\le 1$ by the Euler-Maclaurin formula using up to the fourth Bernoulli number.
This $T^3$-power behavior of the depletion comes from the excitations of Majorana QPs
because $j_s(z,E)$ in the Majorana bound state has a quadratic energy dependence [Fig.~\ref{energyB}(b)].
The observation of the depletion could establish the existence of Majorana QPs.
Since the depletion is proportional to $T^3$, 
the low-temperature depletion of the spin current in the B-phase 
is more gradual than that of the mass current in the A-phase.

\section{Summary and Discussion}

\begin{table}
\begin{center}
\caption{Comparison between the A- and B-phases for a typical $d$-vector,
time reversal symmetry (TRS), a kind of edge current, dispersion of Majorana fermions,
energy dependence of DOS in the bound state, energy dependence of edge current in the bound state,
amplitude of total edge current, low-temperature depletion of edge current,
and energy states where the odd-frequency pair amplitude is related to LDOS.}
\label{tab:summary}
\begin{tabular}{ccc}
\hline\hline
& A-phase & B-phase \\\hline
$d$-vector & $(k_x+ik_y)\bi{z}$ & $k_x\bi{x}+k_y\bi{y}+k_z\bi{z}$ \\
TRS & broken & unbroken \\
kind of edge current & mass current & spin current \\
dispersion & Majorana valley & Majorana cone \\
DOS & constant ($N_0/2$) & $E$-linear \\
edge current & $E$-linear & $E$-quadratic \\
total edge current & $|J|=n\hbar/4$ & $|J_s|=(\kappa/2\pi)(n\hbar/6)$ \\
depletion & $T^2$-power & $T^3$-power \\
odd-frequency & zero energy state & bound states \\
\hline\hline
\end{tabular}
\end{center}
\end{table}

We have found analytic solutions of the quasi-classical Green's function $\hat{g}$ in eq.~\eqref{Green} and eqs.~\eqref{Green0B}-\eqref{GreenzB} for the superfluid $^3$He A- and B-phases with a specular edge, respectively.
These solutions are self-consistent in the weak-coupling limit.
By using the solutions, the DOS including Majorana QPs bound in the edge state has been obtained.
The mass current in the A-phase and the spin current in the B-phase carried by Majorana QPs have also been obtained.
A comparison between the A- and B-phases for the topological features and our main results is summarized in Table~\ref{tab:summary}.

In the A-phase confined to a thin slab, the amplitude of the total edge mass current is $|J|=n\hbar/4$.
The angular momentum by this edge mass current is $L=N\hbar/2$, which corresponds to one of the predictions of the intrinsic angular momentum~\cite{ishikawa:1977,ishikawa:1980}.
This angular momentum can be separated into the contributions of Majorana QPs, $L^{\rm MJ}=N\hbar$, and of the QPs in the continuum state, $L^{\rm cont}=-N\hbar/2$.
Thus, the total angular momentum is given by the cancellation of half of the angular momentum from the Majorana bound state by that from the continuum state.
The cancelled angular momentum $N\hbar/2$ may be related to the DOS of the Majorana QPs $N_0/2$.
The reduction in the angular momentum by temperature behaves as that in $T^2$-power within a low-temperature range by the excitations of Majorana QPs.
Note that, in 3D thick slabs, the angular momentum is reduced by the canting of the $l$-vector at the side edge~\cite{tsutsumi:2011b}.

In the B-phase, the amplitude of the total edge spin current is $|J_s|=(\kappa/2\pi)(n\hbar/6)$.
The numerical coefficient of $n\hbar$ for the total spin current is $2/3$ of that for the total mass current in the A-phase
because the QPs in two of three spin states contribute to the spin current.
The cancellation of a portion of the edge current from the Majorana bound state by that from the continuum state occurs in the B-phase as well as in the A-phase.
QPs contributing to the edge current are not confined in only the vicinity of the Fermi surface, the same as those in the A-phase.
On the other hand, the reduction in the total spin current by temperature behaves as that in $T^3$-power within a low-temperature range owing to the quadratic energy spectrum of the spin current in the Majorana bound state.

Finally, we point out a method of observing Majorana QPs.
The low-temperature depletion of the edge current is due to the excitations of Majorana QPs.
Therefore, the observation of the depletion of the mass current as that in $T^2$-power
or that of the spin current as that in $T^3$-power could establish the existence of Majorana QPs.
We also point out another possibility for experimentally observing Majorana QPs by controlling the surface roughness.
The surface roughness can be controlled by coating the surface with $^4$He atoms,
which has already been proved experimentally~\cite{wada:2008}.
The edge current from the Majorana bound state and that from the continuum state can be suppressed obeying different rates as a function of the surface roughness. 
Particularly in the B-phase, since Majorana QPs are topologically protected, Majorana features are maintained under a certain surface roughness.
In fact, the linear dispersion of the surface Majorana bound state in the superfluid $^3$He B-phase remains intact even at the partially rough surface~\cite{murakawa:2009,murakawa:2011}.
The concrete calculation for quantitatively estimating this change is a future problem.
We note that the edge mass currents in the 2D A-phase for a diffusive edge~\cite{nagato:2011} and a retroreflecting edge~\cite{sauls:2011} have been calculated.

\begin{acknowledgments}

We thank M.~Ichioka, T.~Mizushima, K.~Nagai, S.~Higashitani, and Y.~Nagato for helpful discussions.
Y.T.~acknowledges financial support in the form of the Japan Society for the Promotion of Science Research Fellowships for Young Scientists.
K.M.~acknowledges the support of KAKENHI (No.~21340103).

\end{acknowledgments}

\appendix
\section{Self-Consistency of Solutions}

The self-consistent pair potential $\hat{\Delta }(\bi{r},\bi{k})$ and the anomalous quasi-classical Green's function $\hat{f}(\bi{r},\bi{k},\omega_n)$ satisfy the gap equation
\begin{align}
\hat{\Delta }(\bi{r},\bi{k}) = 
N_0\pi k_BT\sum_{-\omega_c \le \omega_n \le \omega_c}\left\langle V(\bi{k}, \bi{k}') \hat{f}(\bi{r},\bi{k}',\omega_n)\right\rangle_{\bi{k}'},
\label{gap equation}
\end{align}
where $\omega_c$ is the cutoff energy.
The pairing interaction $V(\bi{k}, \bi{k}')=3g_1\bi{k}\cdot\bi{k}'$ 
for Cooper pairs with an orbital angular momentum $l=1$, where $g_1$ is the coupling constant.
The anomalous quasi-classical Green's function is described in spin space as
\begin{align*}
\hat{f}=(i\hat{\bi{\sigma }}\hat{\sigma }_y)\cdot\bi{f}=
\begin{pmatrix}
-f_x+if_y & f_z \\
f_z & f_x+if_y
\end{pmatrix}.
\end{align*}

\subsection{A-phase}

In the A-phase from the solution of the Riccati amplitude in eq.~\eqref{RiccatiA},
the spin component of the anomalous quasi-classical Green's function $f_z=2a/(1+ab)$ is only finite in $\hat{f}$ as
\begin{multline}
f_z(x,\bi{k},\omega_n)=\frac{1}{\sqrt{\omega_n^2+\Delta_A^2\sin^2\theta }}\\
\times\left[d_z(x,\bi{k})-\frac{\Delta_A^2\sin^2\theta\cos^2\phi }
{2(\omega_n+i\Delta_A\sin\theta\sin\phi)}\ {\rm sech}^2\left(\frac{x}{\xi_A}\right)\right].
\label{fzA}
\end{multline}
The second term of $f_z$ contains the odd-frequency pair amplitude~\cite{berezinskii:1974,tanaka:2012}
\begin{widetext}
\begin{align}
f_z^{\rm OF}(x,\bi{k},\omega_n)=-\frac{1}{4}\frac{\Delta_A^2\sin^2\theta\cos^2\phi }{\sqrt{\omega_n^2+\Delta_A^2\sin^2\theta }}
\left[\frac{1}{\omega_n+i\Delta_A\sin\theta\sin\phi }+\frac{1}{\omega_n-i\Delta_A\sin\theta\sin\phi }\right]
\ {\rm sech}^2\left(\frac{x}{\xi_A}\right).
\end{align}
The odd-frequency pair amplitude is related to the LDOS, as pointed out by Higashitani {\it et al}~\cite{higashitani:2012}.
In the Majorana bound state $|E|<\Delta_A\sin\theta$, the real part of the odd-frequency pair amplitude for the energy
\begin{align}
{\rm Re}\left[f_z^{\rm OF}(x,\bi{k},E)\right]\equiv&{\rm Re}\left[f_z^{\rm OF}(x,\bi{k},\omega_n)|_{i\omega_n\rightarrow E+i\eta }\right]\nn\\
=&-\frac{\pi }{4}\frac{\Delta_A^2\sin^2\theta\cos^2\phi }{\sqrt{\Delta_A^2\sin^2\theta -E^2}}
\left[\delta(E-\Delta_A\sin\theta\sin\phi)+\delta(E+\Delta_A\sin\theta\sin\phi)\right]\ {\rm sech}^2\left(\frac{x}{\xi_A}\right),
\end{align}
\end{widetext}
and the angle-resolved LDOS
\begin{multline}
\frac{N(x,\bi{k},E)}{N_0}=\frac{\pi }{2}\frac{\Delta_A^2\sin^2\theta\cos^2\phi }{\sqrt{\Delta_A^2\sin^2\theta -E^2}}\\
\times\delta(E-\Delta_A\sin\theta\sin\phi)\ {\rm sech}^2\left(\frac{x}{\xi_A}\right),
\end{multline}
have the relation
\begin{align}
\frac{N(x,\bi{k},E=0)}{N_0}=\left|{\rm Re}\left[\bi{f}^{\rm OF}(x,\bi{k},E=0)\right]\right|,
\end{align}
in our solution with a specular edge.

At a zero temperature, the $z$-component of the $d$-vector is obtained using the gap equation eq.~\eqref{gap equation} as
\begin{multline}
d_z(x,\bi{k})=g_1N_0\Delta_A\left[k_x\left(\frac{5}{6}+\ln\frac{\omega_c}{\Delta_A}\right)\tanh\left(\frac{x}{\xi_A}\right)\right.\\
\left.+ik_y\left\{\left(\frac{5}{6}+\ln\frac{\omega_c}{\Delta_A}\right)+\frac{1}{4}\ {\rm sech}^2\left(\frac{x}{\xi_A}\right)\right\}\right].
\end{multline}
Since the coupling constant has the relation
\begin{align}
\frac{1}{g_1N_0}=\frac{5}{6}+\ln\frac{\omega_c}{\Delta_A}=\ln\frac{2\omega_c}{\Delta_{\rm BCS}},
\end{align}
where $\Delta_{\rm BCS}$ is the amplitude of the superconducting gap for the conventional $s$-wave superconductivity,
the $z$-component of the $d$-vector is
\begin{multline}
d_z(x,\bi{k})=\Delta_A\left[k_x\tanh\left(\frac{x}{\xi_A}\right)\right.\\
\left.+ik_y\left\{1+\frac{g_1N_0}{4}\ {\rm sech}^2\left(\frac{x}{\xi_A}\right)\right\}\right].
\label{dzA}
\end{multline}
In the weak-coupling limit $\omega_c\rightarrow\infty$ or $g_1N_0\rightarrow 0$,
\begin{align}
d_z(x,\bi{k})=\Delta_A\left[k_x\tanh\left(\frac{x}{\xi_A}\right)+ik_y\right]
\end{align}
in eq.~\eqref{OPA} is a self-consistent pair potential.

\subsection{B-phase}

In the B-phase with the solution of the Riccati amplitude in eq.~\eqref{RiccatiB},
the spin components of the anomalous quasi-classical Green's function $\hat{f}=\hat{M}^{\dagger }\hat{f}'\hat{M}^*$ are described as
\begin{align}
f_x=&\cos(\phi+\theta_L)\ {\rm Im}\left[\frac{2a'}{1+a'b'}\right],\nn\\
f_y=&\sin(\phi+\theta_L)\ {\rm Im}\left[\frac{2a'}{1+a'b'}\right],\
f_z={\rm Re}\left[\frac{2a'}{1+a'b'}\right].
\end{align}
Since
\begin{multline}
\frac{2a'}{1+a'b'}=\frac{1}{\sqrt{\omega_n^2+\Delta_B^2}}\\
\times\left[d'(z,\theta)-\frac{\Delta_B^2\cos^2\theta }
{2(\omega_n+i\Delta_B\sin\theta)}\ {\rm sech}^2\left(\frac{z}{\xi_B}\right)\right],
\end{multline}
by an analogy to the A-phase in eq.~\eqref{fzA}, we obtain
\begin{widetext}
\begin{align}
f_x(z,\bi{k},\omega_n)=&\frac{\cos(\phi+\theta_L)}{\sqrt{\omega_n^2+\Delta_B^2}}
\left[\Delta_B\sin\theta+i\frac{\Delta_B^2\cos^2\theta }{4}
\left(\frac{1}{\omega_n+i\Delta_B\sin\theta }-\frac{1}{\omega_n-i\Delta_B\sin\theta }\right)\ {\rm sech}^2\left(\frac{z}{\xi_B}\right)\right],\\
f_y(z,\bi{k},\omega_n)=&\frac{\sin(\phi+\theta_L)}{\sqrt{\omega_n^2+\Delta_B^2}}
\left[\Delta_B\sin\theta+i\frac{\Delta_B^2\cos^2\theta }{4}
\left(\frac{1}{\omega_n+i\Delta_B\sin\theta }-\frac{1}{\omega_n-i\Delta_B\sin\theta }\right)\ {\rm sech}^2\left(\frac{z}{\xi_B}\right)\right],\\
f_z(z,\bi{k},\omega_n)=&\frac{1}{\sqrt{\omega_n^2+\Delta_B^2}}
\left[\Delta_B\cos\theta\tanh\left(\frac{z}{\xi_B}\right)-\frac{\Delta_B^2\cos^2\theta }{4}
\left(\frac{1}{\omega_n+i\Delta_B\sin\theta }+\frac{1}{\omega_n-i\Delta_B\sin\theta }\right)\ {\rm sech}^2\left(\frac{z}{\xi_B}\right)\right].
\end{align}
The second term of $f_z$ is the odd-frequency pair amplitude.
In the Majorana bound state $|E|<\Delta_B$, the real part of the odd-frequency pair amplitude for the energy
\begin{align}
{\rm Re}\left[f_z^{\rm OF}(z,\bi{k},E)\right]
=-\frac{\pi }{4}\frac{\Delta_B^2\cos^2\theta }{\sqrt{\Delta_B^2-E^2}}
\left[\delta(E-\Delta_B\sin\theta)+\delta(E+\Delta_B\sin\theta)\right]\ {\rm sech}^2\left(\frac{z}{\xi_B}\right),
\end{align}
and the angle-resolved LDOS
\begin{align}
\frac{N(z,\bi{k},E)}{N_0}=\frac{\pi }{4}\frac{\Delta_B^2\cos^2\theta }{\sqrt{\Delta_B^2-E^2}}
\left[\delta(E-\Delta_B\sin\theta)+\delta(E+\Delta_B\sin\theta)\right]\ {\rm sech}^2\left(\frac{z}{\xi_B}\right),
\end{align}
\end{widetext}
have the relation
\begin{align}
\frac{N(z,\bi{k},E)}{N_0}=\left|{\rm Re}\left[\bi{f}^{\rm OF}(z,\bi{k},E)\right]\right|,
\end{align}
in our solution with a specular edge.
The odd-frequency pair amplitude is related to the LDOS not only in the zero-energy state as in the A-phase but also in the Majorana bound state.

At a zero temperature, each component of the $d$-vector is obtained using the gap equation eq.~\eqref{gap equation} as
\begin{align}
d_x(z,\bi{k})=&g_1N_0\Delta_B(k_x\cos\theta_L-k_y\sin\theta_L)\nn\\
&\times\left[\ln\frac{2\omega_c}{\Delta_B}+\frac{1}{6}\ {\rm sech}^2\left(\frac{z}{\xi_B}\right)\right],\\
d_y(z,\bi{k})=&g_1N_0\Delta_B(k_x\sin\theta_L+k_y\cos\theta_L)\nn\\
&\times\left[\ln\frac{2\omega_c}{\Delta_B}+\frac{1}{6}\ {\rm sech}^2\left(\frac{z}{\xi_B}\right)\right],\\
d_z(z,\bi{k})=&g_1N_0\Delta_Bk_z\ln\frac{2\omega_c}{\Delta_B}\tanh\left(\frac{z}{\xi_B}\right).
\end{align}
Since the coupling constant has the relation
\begin{align}
\frac{1}{g_1N_0}=\ln\frac{2\omega_c}{\Delta_B}=\ln\frac{2\omega_c}{\Delta_{\rm BCS}},
\end{align}
each component of the $d$-vector is
\begin{align}
d_x(z,\bi{k})=&\Delta_B(k_x\cos\theta_L-k_y\sin\theta_L)\nn\\
&\times\left[1+\frac{g_1N_0}{6}\ {\rm sech}^2\left(\frac{z}{\xi_B}\right)\right],\\
d_y(z,\bi{k})=&\Delta_B(k_x\sin\theta_L+k_y\cos\theta_L)\nn\\
&\times\left[1+\frac{g_1N_0}{6}\ {\rm sech}^2\left(\frac{z}{\xi_B}\right)\right],\\
d_z(z,\bi{k})=&\Delta_Bk_z\tanh\left(\frac{z}{\xi_B}\right).
\end{align}
Bumps at the edge in the $k_x$- and $k_y$-components are smaller than that in the $k_y$-component for the A-phase according to the coefficients of the squared hyperbolic secant function.
In the weak-coupling limit $\omega_c\rightarrow\infty$ or $g_1N_0\rightarrow 0$,
\begin{align}
\hat{\Delta }(z,\bi{k})=\Delta_B
\begin{pmatrix}
-\sin\theta e^{-i(\phi+\theta_L)} & \cos\theta\tanh(z/\xi_B) \\
\cos\theta\tanh(z/\xi_B) & \sin\theta e^{i(\phi+\theta_L)}
\end{pmatrix}
\end{align}
in eq.~\eqref{OPB} is the self-consistent pair potential.


\end{document}